\documentstyle[twocolumn,twoside,prl,floats,aps,epsfig]{revtex}

\newcommand{\eqa}{\begin{eqnarray}}
\newcommand{\ena}{\end{eqnarray}}
\newcommand{\nn}{\nonumber}
\newcommand{\bra}{\langle}
\newcommand{\ket}{\rangle}
\newcommand{\um}{\frac 12}
\newcommand{\eq}{\begin{equation}}
\newcommand{\en}{\end{equation}}

\tighten
\begin{document}
\draft

\vskip 0.4cm

\title{Simulation of Potts models with real $q$ and no critical slowing down}

\author{Ferdinando Gliozzi}

\address{ Dipartimento di Fisica Teorica
        dell'Universit\`a di Torino, and \\
       I.N.F.N., Sezione di Torino, via P.Giuria 1, 
       I-10125 Torino, Italy\\     
\parbox{14 cm}{\medskip\rm\indent
A Monte Carlo algorithm is proposed to simulate 
ferromagnetic $q$-state Potts model
for any real $q>0$. 
A single update is a random 
sequence of disordering and deterministic moves, one for
each link of the lattice. 
A disordering move attributes a random value to the link,
regardless of the state of the system, while in a deterministic 
move this value is a
state function.
The relative frequency of these moves depends on the two parameters $q$ and
$\beta=\frac 1{kT}$.  The
algorithm is not affected by critical slowing down and the dynamical critical
exponent $z$ is exactly vanishing.
We simulate in this way a $3D$ Potts model in the range $2<q<3$ 
for estimating  the 
critical value $q_c$ where the thermal transition changes from
second-order to first-order, and find $q_c=2.620\pm0.005\;.$
}}
\maketitle
\narrowtext

\section{Introduction}
$Q$-state Potts model \cite{potts} is perhaps one of the simplest,
 non-trivial
 models in 
statistical mechanics. A broad set of techniques has been brought to
 bear on it
in a variety of disciplines and it has been the subject of
 considerable
 theoretical attention over the last two decades (for a review, see \cite{wu}).

This model is theoretically well defined for any real or
 complex value of $q$ \cite{fk}. In particular, the limit $q\to 1^+$ 
corresponds to the random percolation  problem and the limit $q\to 0^+$ has 
a fundamental role in enumerating the spanning trees of a graph \cite{fk}.
Two-dimensional conformal field theory \cite{bpz} suggests  exact
 formulae for the 
critical indices and for other universal quantities as continuous functions 
of $q$ in the range $0<q<4$. Another interesting problem involving 
non-integer $q$ in three-dimensional Potts models is the determination of the 
universal value $q_c$ for which the thermal transition changes from 
second-order to first-order.  A variety of techniques
 have been used \cite{nrs,ks,pk,lk,grt}, 
 which locate $q_c$ in the range $2<q_c<3$. All these methods require 
extrapolations in $q$  because the standard
simulations work only at integer values of $q$.
Reweighting techniques \cite{fmppt,fs} and transfer matrix methods \cite{jc}
 allow to estimate some  thermodynamic functions \cite{lk} in a
wider range of $q$, however there is no way to evaluate correlation
 functions there.

In this paper we remove this limitation by constructing a new Monte Carlo (MC)
algorithm which works for any real $q>0$.
Although the time required for a sweep through the system grows faster
 than its size because at some step of the algorithm nonlocal information is
 required, the
simulations are not affected by a critical slowing down and the dynamical
critical exponent $z$ is exactly zero.
We test the reliability of the method by comparison with some exact results 
for $2D$ Potts model at criticality. We probe its effectiveness by performing
 large scale MC simulations of a three-dimensional Potts model for estimating 
the universal value $q_c$.
\section{the algorithm}
Starting with the Hamiltonian $ H=-\sum_{\bra ij\ket}
\delta_{\sigma_i\sigma_j}$
where the site variable $\sigma_i$ takes the values
$\sigma_i=1,2,\dots,q$, 
with $\bra ij\ket$ ranging over the links of an
arbitrary lattice or graph $\Lambda$, one can write the $q$-state Potts model
partition function $Z=\sum_{\{\sigma\}}e^{-\beta H}$ in the Fortuin 
Kasteleyn (FK) random cluster representation \cite{fk}
\eq
Z=\sum_{G\subseteq\Lambda}W(G)  = \sum_{b,c}\Omega(b,c)\,v^b\,q^c~,
\label{fk}
\en      
\noindent
where $v=e^\beta -1=\frac{p}{1-p}$, the summation is over all spanning 
subgraphs $G\subseteq\Lambda,\;W(G)=v^bq^c$ is their weight, expressed in
terms of the
 number $b$ of edges of $G$, called bonds, and  the number $c$ of
 connected components or FK clusters; $\Omega(b,c)$ is the number of
 subgraphs with $b$ bonds and $c$ clusters. This representation now defines
 a model for any real or complex $q$.

In principle, one could directly use Eq.(\ref{fk}) to define a
Metropolis algorithm
working for positive non integer $q$ \cite{s}, but this is a difficult problem
to simulate because, for each proposed change of a link, the number
$c$ of FK  clusters, a nonlocal property, must be determined.
Large lattices require a huge amount of CPU time.
As a matter of fact, such a method has been applied only to two-dimensional
systems, where special topological relations can be used \cite{s}.

Our strategy is different. We start by considering a useful identity which can
be derived using the methods described in Ref. \cite{cgn}.

Let $l$ be any link of $\Lambda$. Denote by $\{G^+_l\}$ the set of spanning
subgraphs where $l$ is a bond and  by $\{G^-_l\}$ those in which this bond is
missing. 
We have $ Z=Z^+_l+Z^-_l\,$, with $Z_l^\pm=\sum_{G^\pm_l}W(G^\pm_l)$.
Introducing a bond variable $\alpha_l$ equal to $1$ when $l$ is a bond and $0$
otherwise yields
\eq
\bra\alpha_l\ket=\frac{Z^+_l}Z~.
\label{alpha}
\en           
The same quantity can be evaluated in a different way by addition of a bond to
each graph of type $G^-_l$.
There are two kinds of missing bonds. Those joining two different clusters,
called potential bridges, are picked out by a variable $\beta_l$ which takes
the value $1$ only on them and is zero otherwise;
their addition lowers the number $c$ of FK clusters. We have
\eq 
\beta_l=1\Rightarrow ~~W(G^-_l)\frac vq=W(G^+_l)\;\;.
\label{beta}
\en           
The remaining missing bonds,  described by a similar variable
$\gamma_l$, join two sites of the same cluster; their addition keeps
$c$ invariant, then
\eq
\gamma_l=1\Rightarrow~~ W(G^-_l)v=W(G^+_l)\;\;.
\label{gamma}
\en        
Combining Eq.s (\ref{alpha}), (\ref{beta}) and (\ref{gamma}) yields
$ \bra \alpha_l\ket=\frac vq\bra\beta_l\ket + v\bra\gamma_l\ket$
which is the wanted identity.
Since of course $\alpha_l+\beta_l+\gamma_l=1$, we can rewrite it as
\eq
\bra\alpha_l\ket=p\,\bra\alpha_l\ket+\frac pq\,\bra\beta_l\ket+
p\,\bra\gamma_l\ket\;\;, 
\label{id}
\en  
where the weighting factors can now  be interpreted in 
terms of probabilities.
The idea is now to regard this identity as the limit of a recursion relation of
the type
\eq
\pi^{(n+1)}_l=p\,\alpha^{(n)}_l+\frac pq\,\beta^{(n)}_l+p\,\gamma^{(n)}_l\;\;,
\label{recur}
\en
where $\pi^{(n+1)}_l$ is the probability of having a bond on the link
 $l$ in the configuration $G^{(n+1)}$. It is expressed as a state function 
 $(\alpha_l,\beta_l,\,{\rm or}\,\gamma_l)$ of the same link in the 
$G^{(n)}$ configuration. This generates a Markov process
$ \cdot\cdot\to G^{(n)}\to G^{(n+1)}\to\cdot\cdot$ 
where the equilibrium distribution yields Eq. (\ref{id}).
This stochastic chain fullfills two important conditions: 
$i)$ there is a non-zero probability of going from any configuration
to any configuration in a single sweep through $\Lambda$, 
$ii)$ the equilibrium distribution maps to itself 
as Eq.(\ref{id}) is kept invariant by the process. One can then argue
that detailed balance is satisfied. 

To see it directly, assume for
instance that  in the $n^{th}$ configuration $l$ is a 
potential bridge 
 $(\beta^{(n)}_l=1\;,\;\;G^{(n)}=G^-_l)$ which is promoted to a bond in the 
$(n+1)^{th}$ configuration $(\alpha^{(n+1)}_l=1\;,\;\;G^{(n+1)}=G^+_l)$.
The transition rate is $P(G^-_l\to G^+_l)=\frac pq$.
Conversely Eq.(\ref{recur}) yields $P(G^+_l\to G^-_l)=1-p$.
Then, according to Eq.s (\ref{fk}) and (\ref{gamma}),
\eq
\frac {P\left(G^-_l\to G^+_l\right)}{P\left(G^+_l\to G^-_l\right)}=
\frac {W\left(G^+_l\right)}{W\left(G^-_l\right)}
\en
as detailed balance requires. The same conclusion can be reached in
all the other cases.

A straightforward, preliminary, implementation of the recursion
relation (\ref{recur}) is the following: 
$i)$ go over each link 
$l\in\Lambda$ of the configuration $G^{(n)}$ and generate a
pseudo-random number $X_l$ uniformly
distributed from $0$ to $1$. $ii)$ Create a bond on $l$ only 
in the following two cases: 
a) $X_l<p$ \underline{and} $l$ is a bond  $(\alpha_l=1)$ or a missing bond
joining two sites of the same FK cluster $(\gamma_l=1)$
; b)  $X_l<\frac pq$ \underline{and} $l$
is a potential bridge $(\beta_l=1)$. 
This generates uniquely the configuration $G^{(n+1)}$.

Let $q> 1$ for definiteness. It is worth noting that when $X_l<\frac
pq$ the algorithm adds a bond to $l$ regardless of which configuration 
$G^{(n)}$ we are dealing with.
Similarly, when $X_l>p$ no bond is added. In the remaining cases 
$(\frac pq\le X_l\le p)$. The value attributed to $l$ 
(bond or no bond) is unambiguously determined by $G^{(n)}$. 

Inspecting  all the cases leads  to the
following better implementation of the algorithm:

Step 1: Pick a link $l\in\Lambda$ and generate a pseudo-random 
number $0\le X_l\le1$.

Step 2: Update the link according to the following scheme 
\begin{eqnarray}
{\rm ~~~ move}&&\hskip -.5 cm{\rm current~ state} ~~{\rm new~ state}\nn\\
 {\rm a)}~~X_l&<~\frac{p}q~~&~{\rm any}~~~~~~~~~~{\rm bond}\nn\\
 {\rm b)}~~X_l&>~p~~~\hfill&~{\rm any}~~~~~~~~~~{\rm no~bond}\nn\\
{\rm c)}~~\frac pq~&\le  X_l\le p~~~\hfill&\cases{\alpha_l=1&~~bond\cr
\beta_l=1&~~no bond\cr\gamma_l=1&~~bond\cr} \nn\\
\nn
\end{eqnarray}

Step 3: Return to step 1.

The first two moves do not need any information on the state of the 
system: they just disorder it. The last one is a purely deterministic 
move; its only effect is to put a bond whenever a link joins two sites
of the same cluster. It requires 
distinguishing between the two kinds of missing bonds 
$(\beta_l=1~ {\rm or}~ \gamma_l=1)$.
One can infer this nonlocal property by identifying the connected 
components of the configuration, like in the Swendsen-Wang (SW)
algorithm \cite{sw}. 
This cluster reconstruction is  time demanding, however it
gives a complete information on the state of the missing bonds of the 
whole lattice.
As the update proceeds through the lattice this amount of information
is progressively
lost because of  disordering moves (the deterministic moves never
change $c$). We may partly keep track of the cluster
structure by relabelling the cluster indices whenever a disordering move 
creates a bond
between two of them. Cluster reconstruction is truly necessary only
when
 a deterministic move touches a missing bond  of a putative 
single cluster where
some bond has been erased by  previous disordering moves.

Because of non locality, the number of operations involved every MC step is
$\propto N^\alpha$, where $N$ is the number of links and $1<\alpha\le 2$.
The efficiency of the algorithm depends 
crucially on the actual number of cluster
reconstructions per sweep. In our $3D$ simulations reported below the 
fraction
of links requiring cluster reconstruction was about $3\%$ with a
decreasing trend for larger lattices.
\section{correlation times}
An unusual  feature of the  described algorithm is the presence of randomly 
distributed disordering moves.  The mean number of links subjected to  
disordering moves in a single sweep is $Np_r$ with
$ p_r=1+\frac pq-p~. $
For instance, in the Ising model $(q=2)$ at criticality 
 more than $70\%$ of the links are disordered every sweep. 
It is now easy to find an upper bound for the mean number $\tau$ of MC steps 
needed to generate effectively independent configurations. 
After $n$ sweeps the mean number of  links which do not have yet undergone 
 a disordering move is $N(1-p_r)^n$. When this number is of the order of 1 
all the links have been touched by a disordering move and the 
 upper bound $\tau_o\ge\tau$ is given by the obvious relation 
$N(1- p_r)^{\tau_o}\sim 1$, i.e.
$\tau_o=-\frac{\log N}{\log(1- p_r)}$.  
Thus the dynamical exponent $z$ is 0, as critical 
slowing down manifests itself by the power law  
$\tau\propto N^z$ at the critical temperature 
where a second order phase transition occurs \cite{wang}.
A numerical estimate of the decorrelation time of the dynamics of the
new algorithm  for the critical Ising model on a square lattice is reported
in Tab. \ref{tab1}. Note that the actual value of $\tau$ does not
saturate the upper bound and is much smaller than the analogous
quantity of the SW algorithm.
\begin{table}
\caption[] {The decorrelation time $\tau$  
of the new algorithm for the critical 2D Ising model  for 
different linear lattice
sizes $L$  is compared with the same quantity of the  SW algorithm 
$\tau_{SW}$ and with the upper bound $\tau_o$. The definition of 
$\tau$ and the $\tau_{SW}$ data are taken from Ref. \cite{wang}. }
\begin{tabular}{cccc}
$L$&$\tau$&$\tau_{SW}$&$\tau_o$\\
\hline
8&2.65(3)&5.17696(32)&3.3869\\
 16&   3.16(5)&6.5165(12)&4.5158\\
 32&   3.69(6)&8.0610(18)&5.6448\\
 64&   4.3(1)&9.794(4)&   6.7737\\
\end{tabular}
\label{tab1}
\end{table}

The new algorithm proves also useful in fighting against another 
  dynamical problem which one deals with in the case of 
first-order transitions, namely the exponentially fast suppression of
 the tunnelling  between metastable states with increasing lattice
 size.

To reduce this type of slowing down the multi-canonical MC algorithm
 has been proposed \cite{bn}; also the method of simulated tempering
 \cite{mp} proves useful \cite{kw}. In a few numerical tests for
 two-dimensional models with $q=7$ and $20\ge L\ge100$ we found that
 the tunnelling time of the canonical algorithm described in the
 present paper grows with the system size $V$ as $\tau_t\propto
 V^\alpha$ with $\alpha=1.03\pm0.03$, like in an optimal variant \cite{jk}
 of the multi-canonical method, but with a smaller proportionality
 factor.

The reason of this performance is that the random moves accelerate the
 tunnelling between order and disorder. The drawback is that the new
 algorithm is non-local, so the CPU time grows like $V^b$ with $b>1$;
 for instance in the present case we found $b\sim 1.85$. Thus the new
 algorithm cannot certainly be recommended for integer $q$, although at a
 first-order transition it  performs much better than any local 
canonical algorithm.

\section{Simulations}
As a first, simple, application of the new algorithm we tested
the reliability of our code by checking a percolation property of Potts model 
on a square lattice which is supposed  to be exact in the range
$0<q<4$, namely  that the mean frequency of active bonds
$\bra\alpha_l\ket$ at criticality (corresponding to $v=\sqrt{q}$ in 
Eq.(\ref{fk})) should coincide, in the thermodynamic limit, with the 
random percolation value, i. e. $\bra\alpha_l\ket=\um$,  irrespective 
of the value of $q$\cite{wu}.
We simulated critical Potts models in a $128\times128$ square lattice 
with $q$ ranging from $1.5$ to $3.5$. In all the cases the mean number
of bonds was compatible with the exact result. Finite size effects of
this observable, which are visible on smaller lattices, allow to
evaluate the critical thermal exponent $\nu$ as a function of $q$.  
This could be used to check a conjectural formula suggested by the
2D conformal field theory \cite{bpz}. We plan to study this 
problem in a future publication.

\begin{figure}[t]
\label{f1}
\begin{center}
\mbox{~\epsfig{file=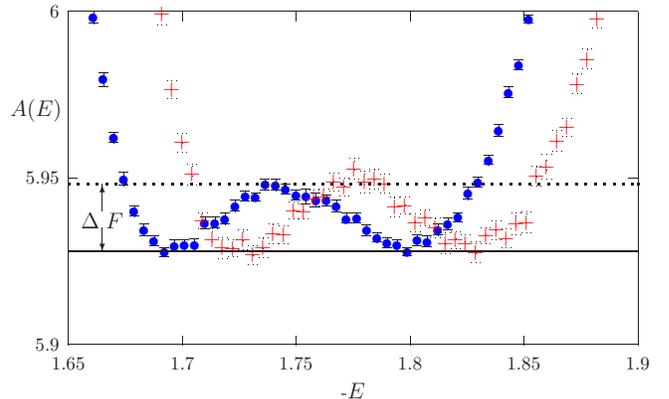,width=8.5cm}}
\caption{ 
Plot of $A(E,L)$ resulting from a simulation of a $3D$ Potts model at
$q=2.75$ with 3.3 $10^7$ MCS for $L=14$ (full circles) compared with the
extrapolation at the same value of $q$ of an actual SW simulation of
4.1 $10^7$ MCS at $q=3$ (crosses). The latter data are shifted to the 
right for clarity.}  
\end{center}
\vspace{-0.75cm}
\end{figure}

 The new algorithm allows us to deal with an important issue of
 three-dimensional Potts model, namely the estimate of the tricritical
 point $q_c$ in the range $2<q_c<3$, where  the thermal transition
 changes from second-order for $q\le q_c$ to  first-order for
 $q>q_c$. Many different  techniques
 have been used  to locate this point \cite{nrs,ks,pk,lk,grt}.  
 We applied a method very similar to that described 
 by Lee and Kosterlitz \cite{lk} by computing the double
 histogram $N(b,c)$ of
 bond and cluster number distribution in a cubic lattice of volume
 $L^3$ at a given $\beta$ and $q$ and then extrapolating the data to
 nearby values. Using Eq.(\ref{fk}) we can write 
 \eq    
 N(b,c;\beta,q,L)={\cal N}\,\Omega(b,c)\,\frac{v^b q^c}Z,
 \label{histo}
 \en
\noindent where ${\cal N}$ is the number of MC sweeps.
We can trade the number of bonds $b$ for the energy per site $E$ using
the relation $E=-b\frac{v+1}{v\,L^3}$.
Near a first-order transition the histogram $P(E)=\sum_cN(b,c)/{\cal
  N}$ has a characteristic
double peak structure corresponding to the ordered and the disordered
phase. A suitable reweighting through Eq.(\ref{histo}) of the energy 
distribution yields the value $\beta_c(L,q)$ where the two peaks 
at $E_1(\beta,L)$ and $E_2(\beta,L)$  are of equal
height. A typical plot of the quantity
$A(E,q;\beta_c,L)=-\sum_c\ln(N(b,c)/{\cal N})$ is shown in
Fig.1. 
A useful
estimator of the interface free energy between the ordered and the
disordered phase \cite{b} is given by
\eq
\Delta F(q,L)=A(E_m,q;\beta_c,L)-A(E_1,q;\beta_c,L)~,
\label{df}
\en
where $E_m$ is the local maximum which separates the two dips at $E_1$
and $E_2$ (see Fig.1).  At a first-order transition $\Delta
F(L)$  increases 
monotonically with $L$ and is expected to vanish at the tricritical
point. Extrapolating the numerical data both in $\beta$ and $q$ one
may locate this point. The region of reliable extrapolation \cite{fs} is 
$O(1/L^{3})$ for both $\beta$ and $q$. This does not cause a problem for 
$\beta$, since it can be adjusted 
continuously, but $q$ cannot in standard simulations, being by
necessity an integer value. Actually Lee and Kosterlitz  performed
their 
simulations at $q=3$ and found that the extrapolated data become too noisy for 
$\vert\delta q\vert>0.3\;$ \cite{not}. In our case we can directly
evaluate the range of reliable extrapolations. Indeed  
the main advantage of the algorithm described in this paper is that now
also $q$ can  be adjusted continuously.
 
Our simulations were performed on three different lattices as listed
in Tab.\ref{tab2}.
\begin{table}
\caption[]{The simulations were performed on cubic lattices of side
  $L$ at the values of $q$ and $\beta$ listed below. $MCS$ is the number of 
Monte Carlo steps considered.}
\begin{tabular}{cccc}
$L$&$q$&$\beta$&$MCS$\\
\hline
12&2.70&0.52270&$3.0~10^7$\\
13&2.70&0.52270&$3.0~10^7$\\
14&2.75&0.52721&$3.3 ~10^7$\\
\end{tabular}
\label{tab2}
\end{table}
The statistics is good since in all the cases the mean flipping time 
between coexisting states was no larger than thirty
MC steps. The errors were calculated by gathering the histogram $N(b,c)$
every $10^6$ MC steps and then performing a standard analysis. 
\begin{figure}[t]
\label{f2}
\begin{center}
\mbox{~\epsfig{file=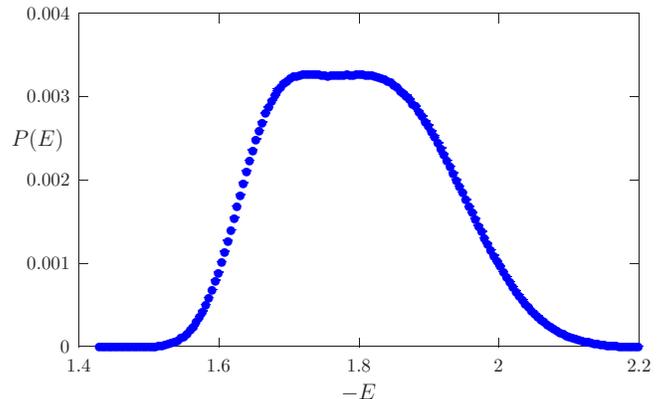,width=8.5cm}}
\caption{ Energy histogram at $\beta_c(L,q)$ obtained by a simulation at
  $q=2.7$ and $L=13$. 
}  
\end{center}
\vspace{-0.75cm}
\end{figure}

In all the cases the energy histogram showed a double peak structure,
providing a direct evidence of the first-order nature of the
transition for these values of $q$ (see Fig.2). 
This yields the upper bound
$q_c<2.7$. In shorter simulations at $q=2.6$ we found no trace of a
double peak structure. This suggests $q_c>2.6$.
Using  the reweighting
method we estimated  the values $\beta_c(L,q)$ where the two peaks are 
of equal height for each  $L$ and for few values of $q$ near 
$q=2.7$ and  the corresponding values of $\Delta F$. The results are
reported in Fig.3. 

A further
reweighting up to $q=3$ allowed to compare the extrapolated data with  
those coming from a similar extrapolation of standard SW simulations 
at $q=3$. This comparison showed that the range of reliable
extrapolations is $\vert\delta q\vert<0.25$. 
It has to be noted that we could not use for this 
comparison the high precision data of Ref. \cite{jv}, because the
energy distribution in terms of spin variables used there 
does not coincide with that expressed in terms of bond variables
used by necessity in the present approach. In particular the 
$\beta_c(L,q)$'s are shifted and our $\Delta F(q,L)$ is always 
smaller.

Simple finite size scaling considerations
suggest \cite{lk} that near $q_c$ the interface free energy has the
simple form $\Delta F(q,L)\sim (q-q_c)^2L^{a}$ which fits very well to
our data (see Fig.3). To within our numerical accuracy $a=4.8\pm0.1$ 
and $q_c=2.620\pm0.005$. This agrees with the value $q_c=2.55\pm0.12$ 
obtained in the large $q$ expansion of the latent heat \cite{ks}.
Lee and Kosterlitz \cite{lk} extrapolating $q=3$ data found  a smaller 
value, $q_c=2.45\pm.0.10$. The difference could be due to the fact
that extrapolations with $\vert\delta q\vert>0.25$ gives an
overestimate of $\Delta F$ (this is already visible in Fig.1).
Other approximate methods give even smaller values: real space
renormalization group methods \cite{nrs} yield $q_c\sim 2.2$ while
an Ornstein-Zernike approximation \cite{grt} gives $q_c\sim 2.15$.

\begin{figure}
\label{f3}
\begin{center}
\mbox{~\epsfig{file=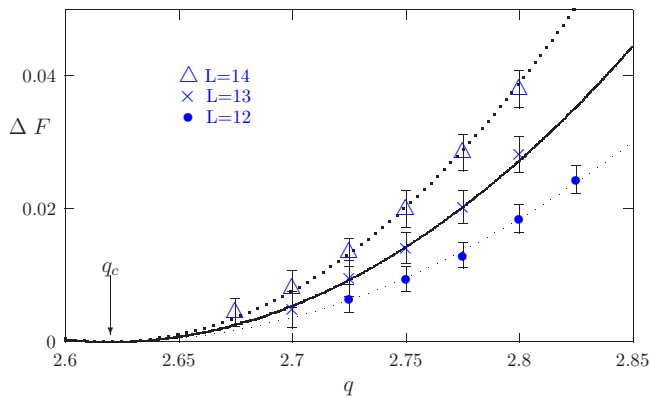,width=8.5cm}}
\caption{ Plot of $\Delta F(q,L)$ near $q=2.7~$. 
}  
\end{center}
\vspace{-0.75cm}
\end{figure}

\section{CONCLUSIONS}
This work provides a new MC algorithm to simulate ferromagnetic $q-$
state Potts model which has two very unusual features: it works for any
real $q>0$ and does not suffer of any critical
slowing down. The former property is an obvious consequence of the fact
that it is based on the Fortuin Kasteleyn random cluster
representation, where $q$ acts as a continuous parameter. The latter
is more tricky and is due to the implementation of the algorithm with 
a random sequence of disordering moves, randomly distributed over the
lattice. There is no reason to believe that this disordering mechanism is
specific to Potts model and it would be very interesting to try to
implement it in other, more general MC methods. A drawback of the new 
algorithm is that it is non-local, so the CPU time of a single sweep
grows with the volume $V$ as $V^b$ with $1<b<2$, thus it is
not recommended for integer $q$, where the SW algorithm works 
with $b=1$. Actually at a first-order transition the new algorithm
performs better than the SW method, but there the multi canonical 
MC algorithms are more suitable.

It is straightforward to extend the new algorithm in order to take
into account quenched bond randomness, provided that all the couplings
are ferromagnetic. On the contrary, generalising to systems with
frustrations seems a rather difficult task, because it is not obvious
how to define in this case the FK clusters for non-integer $q$ \cite{cgn}.

We used such an algorithm to study the region $2<q<3$ of
a three-dimensional Potts model in order to  estimate the critical value $q_c$
for which the thermal transition changes from second to
first-order. We obtain a rather precise estimate compared to other
methods,\cite{nrs,ks,pk,lk,grt}, the reason being that all the
other methods are based on  extrapolations from integer values of
$q$, while the new algorithm simulates the system at nearby values of $q_c$.
\vskip .2 cm

The author would like to thank M. Caselle and A. Coniglio for helpful 
discussions.


\begin{references}
\bibitem{not} Recently it has been suggested  that
a refined form \cite{wjh} of the multi-histogram technique \cite{fs2} could
allow for the extrapolation of thermal averages in a broad range of
$q$.
\end{references}
\end{document}